\documentclass[10pt,twocolumn]{elsart}
\usepackage{natbib}
\usepackage[dvips]{graphicx}
\usepackage{amssymb}
\usepackage{psfrag}
\journal{New Astronomy}

\def\astrobj#1{#1}

\begin{document}

\begin{frontmatter}
\title{Disk/corona model of the mean spectra of radio-quiet and radio-loud QSO}
\author[CAMK]{M.~Niko\l{}ajuk \corauthref{cor}},
\corauth[cor]{Corresponding author.}
\ead{mark@camk.edu.pl}
\author[CAMK]{B.~Cze\-rny}
\ead{bcz@camk.edu.pl}

\address[CAMK]{Nicolaus Copernicus Astronomical Center,
Polish Academy of Sciences, Bartycka~18, 00-716~Warsaw, Poland}

\begin{abstract}
We test the disk/corona model of Janiuk et al.~(2000) against the
composite mean optical/UV/soft X-ray spectrum of radio loud and radio
quiet quasars from \citet{Laor97} which applies to faint quasars.  The
model well represents the optical/UV continuum if the hardening of the
locally emitted spectrum is included, with the color to effective
temperature ratio $f \sim 2$ in the inner 20-30 Schwarzschild
radii. Comptonization seen in the soft X-ray band is well explained by
the adopted corona model in the case of radio loud objects. However,
this Comptonization is much stronger in radio quiet objects and
additional {\it ad hoc} assumed Comptonizing medium must be
present. We speculate that perhaps in both types of quasars there is a
strong outflow of the hot plasma; this plasma is collimated in radio
loud objects but not collimated in the radio quiet objects so it is
present along the line of sight, serving as the required Comptonizing
medium. Hard X-ray power law is not explained by the model and it may
come from non-thermal component of electron plasma, as it is the case
for galactic black holes in their soft states.
\end{abstract}

\end{frontmatter}

\section{Introduction}
The standard model of central engine of active galactic nuclei (AGN)
consists of a massive black hole, an accretion disk, and a distant
molecular torus. A major problem, however, is the location of the
source of X-ray emission not expected from this simple picture. X-ray
emission carries a significant fraction of the bolometric
luminosity. In Seyfert 1 galaxies it is mostly in the form of a hard
X-ray power law component with the high energy cut-off at about 100
keV. In Narrow Line Seyfert 1 galaxies and quasars the hard X-ray
power law contains much lower fraction of the bolometric luminosity
but there is a second component in those sources - large soft X-ray
excess - which carries large fraction of the luminosity. This X-ray
emission comes from a relatively compact region located not too far
from the accretion disk, as suggested by the fast variability in X-ray
band and the presence of X-ray reflection component. However, even the
extensive monitoring campaigns did not solve yet the problem of exact
geometry of this region and, subsequently, of the physical process
leading to the formation of the hot plasma.

In the present paper we consider a particular model of the origin of
this X-ray emission which is based on physical grounds. It is a model
of a hot dissipative corona above an accretion disk, as described by
Janiuk \& Czerny (2000). This model is based on assumptions that the
corona is in hydrostatic and radiative equilibrium with the disk, the
rate of energy dissipation in the corona as well as in the disk is
proportional to the pressure (total pressure in the disk and gas
pressure in the corona) and the spontaneous division of the flow into
disk and corona reflects the thermal instability within the irradiated
gas, as discussed by \citet{KMT81}. This model predicts the overall
shape of the spectrum optical/UV/X-ray from an accretion flow as 
a function of the mass of the black hole, an accretion rate and the
viscosity parameter $\alpha$. We introduce some modifications to the
model and we compare it with the observational data.

One of the source of difficulties in modeling AGN is the fact that
significant fraction of the bolometric luminosity is emitted in the
unobserved region between far UV and soft X-rays so the bolometric
luminosity and the spectral shape of the principal component - Big
Blue Bump - are poorly determined for a single object. Therefore, in
the present paper we decided to test the model against composite
quasar spectra of \citet{Laor97} which practically bridge the
unobserved gap due to combining the spectra coming from objects at
various redshifts.

In Section~2 we shortly describe the composite and a theoretical model.
Section~3 describes our results. We discuss the results in Section~4.
The conclusions are given in Section 5.
\section{Method}
\subsection{Composite spectrum}

The mean spectra of radio quiet and radio loud quasars were shown by
\citet{Laor97}. They cover a broad frequency range $log\,\nu \simeq
14.3-18.5$, i.e. they extend from 15 000 \AA~ up to 10 keV.

The mean X-ray spectrum contains the spectra of objects from the
Bright Quasar Survey \citep{SG83} with redshift $\mathtt{z} \leq 0.40$ and
column density $N_{H I}^{Gal} < 1.9 \times 10^{20} \
\mathrm{cm}^{-2}$.  It was obtained adopting the following
cosmological parameters: the Hubble constant $H_0$ equal $50\
\mathrm{km\: s}^{-1}\, \mathrm{Mpc}^{-1}$, the deceleration parameter
$q=0.5$, and the cosmological constant $\Lambda$ equal to $=0.0$.
These low redshift objects determine the mean spectrum above $\log\,
\nu \approx 16.7$ (200 eV). Below it we have the unobserved range due
to the Galactic absorption which extends from $log\, \nu
\simeq 15$ to $log\, \nu \simeq 16.7$. This range, however, is
partially covered from $2000$ \AA~ to $350$ \AA~ ($log\, \nu \sim
15.2-16.0$) by spectrum from \citet{Zheng97}.  This result is based on
observations of quasars at high redshifts ($z >0.33$).

Final spectrum still contains a relatively short uncovered
range. However, the simple extrapolation of the spectra from both
sides roughly coincide which strongly suggest that the resulting
spectra are a good representation of the broad band spectra of bright
radio quiet and radio loud AGN.

\subsection{Disk/corona/extended~medium structure}
\label{sect:model}

We consider a model of stationary accretion flow onto a Schwarzschild,
massive black hole which consists of an accretion disk, a hot corona
and (optionally) a spherically symmetric extended hot medium. The
accretion is ultimately responsible for the emission of the radiation,
and the energy is dissipated partially in the disk and partially in
the corona.  Therefore, the accretion rate, $\dot M$, together with
the mass of the central black hole, $M$, are the basic parameters of
our model \citep[see also][]{WCZ97, JC2000}. 
The total flux generated in the corona, $F_{c}$, and in the disk,
$F_{d}$, is determined by the standard formula \citep[see e.g.][]{KFM98}
\begin{equation}
F_{d}+F_{c}= \frac{3GM\dot M}{8\pi r^{3}}\mathcal{G}(r) \ ,
\end{equation}
where $\mathcal{G}(r)$ represents the Newtonian boundary condition at
the marginally stable orbit
\begin{equation}
\mathcal{G}(r)=1-\sqrt{\frac{3\mathrm{R_{Schw}}}{r}} \ .
\end{equation}

\subsubsection{Local disk spectrum}

We assume that the disk is relatively cold, geometrically thin and
optically thick \citep[i.~e. $\alpha$--disk theory,][]{ShaSuny73}.  We
approximate the distribution of the angular momentum in the disk and
the corona by Keplerian distribution
\begin{equation}
\Omega_{K} = \sqrt{\frac{GM}{r^{3}}} \ ,
\label{Omega}
\end{equation}
where $G$ is the gravitational constant, $M$ -- the mass of the black hole
and $r$ is radial radius of the object. The disk is the source of 
the soft photons, $F_{soft}$, which are Comptonized locally in the corona
and afterwards, at a greater distance, by the spherical cloud. 
The flux $F_{soft}$ consists of a flux $F_d$ generated in the disk at any 
given radius is denoted by $F_{d}$ and of a fraction of the flux $F_c$
generated in the corona and subsequently intercepted and thermalised 
in the disk. This flux determines the effective temperature of the disk
at a given radius
\begin{equation}
F_{soft} = \sigma T_{eff}^4 \ .
\end{equation}

The spectral shape of this radiation is sometimes approximated as a
blackbody. However, when an electron scattering dominate significantly
over free-free and bound-free absorption this assumption is not
justified. If the number of scattering events is large the saturated
Comptonization effect is the most important and it shifts
systematically the energy of photons while still preserving a 'black
body' spectral shape.  \citet{ShiTa95} showed that this is
an important effect in AGN accretion disks.
 
We introduce the spectral hardening factor $f$ in such way that the
flux can be described by the equation \citep{ShiTa95}
\begin{equation}
F^{soft}_{\nu} = \frac{1}{f^{4}}\pi B_{\nu}(fT_{eff}) \ ,
\label{eq:fnusoft}
\end{equation}
where $T_{eff}$ is the effective temperature, $T_{col}=fT_{eff}$ is  
the color temperature, and $B_{\nu}$ is the Planck function of blackbody.

The value of $f$ should come in principle from the radiative transfer
computations. However, the results of Shimura \& Takahara (1995) were
obtained assuming pure hydrogen and helium atmosphere while heavy
elements (particularly iron) are very important. Other codes
\citep[e.g.][]{MFR2000,Sob99,MaR2000}
also contain several approximations, mostly
concerning the description of bound-free transitions and lines of
heavy elements which are essential.  It is therefore more convenient
to parameterize this effect, compare the results to the data and 'a
posteriori' conclude which theoretical predictions are most
appropriate.

This effect is particularly important in the innermost part of the disk where
the ionization is the highest but it should become negligible further out.
Therefore we assume that the 
spectral hardening factor $f$ is the monotically decreasing function.
It decreases from the value $f_{max}$ at the marginally stable orbit radius 
to the value $f = 1$ further away, at $r$ larger than $r_{col}$.  
We use basically a linear function but results obtained with other
functional shapes are similar (see Sect. 3).

\subsubsection{Local corona}
The corona is hot, optically thin and two-temperature (i.~e. the ions temperature $T_{i}$ is much greater than the electrons 
temperature $T_{e}$, as discussed by \citet{ShaLigE76}). The two-temperature
corona model was considered by a number of authors (\citet{HaarMar91,WCZ97,Esin97}). 
Here we follow the formulation of \citet{JC2000}.
 
We assume that the coronal plasma is an isothermal medium in the vertical 
coordinate, $z$, but non-isothermal toward the radial coordinate, $r$ (i.~e. 
$\frac{\partial T}{\partial z} = 0$ and $\frac{\partial T}{\partial r} \neq 0$).
\begin{figure}[t]
\centerline{
\includegraphics[width=0.45\textwidth, angle=0]{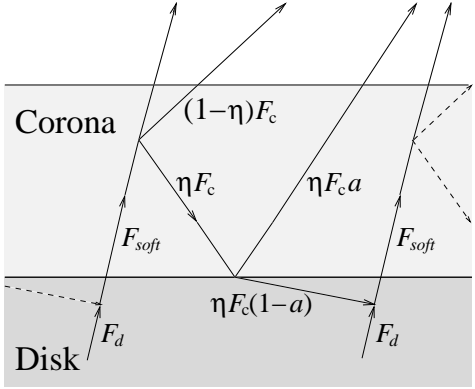}}
\label{strum.eps}
\caption{\small A schematic diagram of radiation fluxes distribution in 
the two-temperature corona model. $F_{d}$ and $F_{c}$ are the fluxes 
produced, respectively, in the disk and in the corona. 
$F_{soft}=F_{d}+\eta F_{c}(1-a)$ is the flux of soft photons, which are 
Comptonized without disk. $\eta$ is a fraction of coronal photons directed
towards the disk, $a$ is a disk albedo. 
}
\end{figure}
The flux of the soft photons, $F_{soft}$, produced in the disk is
partly Comptonized within the corona and it is partly transmitted
throughout. The scattered Comptonized coronal flux, $F_c$, is
partially directed towards the disk, $\eta F_{c}$, and partially
directed outwards to an observer, $(1-\eta) F_{c}$, as a hard X-ray
and gamma-ray radiation (see Fig.~1). X-ray photons irradiating the
disk are partially absorbed and thermalised, contributing to
$F_{soft}$, and partially reflected by the disk surface, depending on
the value of the X-ray albedo, $a$.  We assume in our calculations
that $\eta = 0.5$ and $a = 0.2$. Such a formulation of the radiative
disk/corona coupling was introduced by \citet{HaarMar91}.

The local flux $F_{c}$ is generated by a heating mainly of the ions, 
because a gravitational energy by accreting gas is transported
directly to this particles \citep{NMQ98}. In this case the flux $F_{c}$ is 
proportioned to $\alpha$-viscosity introduced by \citet{ShaSuny73}
and the total pressure $P$.
\begin{equation}
F_{c}= \frac{3}{2} \Omega_{K} \alpha \int_{z_0}^{\infty} P(z) \, dz  \ ,
\end{equation}
where $z_0$ is the vertical coordinate at the basis of the corona.
In our calculate we neglect the coronal radiation pressure and
the magnetic pressure, 
and we employ the results of the vertical
integration obtained by \citet[Appendix~D]{WCZ97}.  
At present 
\begin{equation}
F_c = \frac{3}{2} \Omega_K \alpha P_0 H \sqrt{\frac{\pi}{2}} \ ,
\end{equation}
where $P_0 \equiv P(z_0)$ is the gas pressure at the basis of the corona,
$H$ is the pressure scale height of the corona given by the ions temperature,
$T_i$, under the assumption of the hydrostatic equilibrium
\begin{equation}
H = \frac{1}{\Omega_K} \sqrt{\frac{kT_i}{m_H}} \ .
\end{equation}

Such a formulation is independent from the physical mechanism of the corona 
heating since the scaling with pressure, $H$, may correspond 
either to the accretion heating by viscosity \citep{ZCC95,WCZ97} 
or to the magnetic heating \citep[e.g.][]{SZ94}.

The energy of ions in the corona is immediately  transported to electrons 
through the electron-ion Coulomb interaction. The ions are cooled and 
this cooling is described by the equation \citep{ShaLigE76}
\begin{equation}
F_c=\frac{3}{2} \frac{k(T_i -T_e)}{m_H} 
\left[1 + \sqrt{\frac{4kT_e}{m_e c^2}} \right] \nu_{ei}\rho_0 H 
\frac{\sqrt{\pi}}{2} \ ,
\label{4.18}
\end{equation} 
where $\nu_{ei} = 2.44 \times 10^{21} \rho_0 T_e^{-1.5} \ln \Lambda \: [\mathrm{s}^{-1}]$
is the electron--ion coupling rate and $\ln \Lambda \approx 20$ is the
Coulomb log. Here the density $\rho_0$ is related to the corona thickness 
and its optical depth $\tau_{es}$.

The heated electrons deliver the energy subsequently to photons
in the Inverse Compton process. In this way the part
of observe/UV/soft X-ray photons emerge to a observator as  
a hard X-ray radiation. We describe this process as
\begin{equation}
F_c= A(\tau_{es}, T_e, T_s) F_{soft} \ ,
\end{equation}
where $A(\tau_{es}, T_e, T_s)$ is the Compton amplification factor
and $T_s$ is the temperature of disk produced the local $F_{soft}$.
We interpolate  the factor $A$ from the tables prepared in advance from
Monte Carlo computations. 

The vertical division of the medium into hot corona and relatively
cold disk at every radius should not be arbitrary. Indeed, such 
a division results naturally from the criterion of thermal instability
in a irradiated medium studied by \citet{KMT81} which naturally develops
in the surface layers of an irradiated accretion disk 
in hydrostatic equilibrium \citep{RCz96}.

The ionization stage of the medium is conveniently expressed
through a ionization parameter $\Xi$. It is defined as a ratio of
the ionization radiation pressure to the gas pressure
\begin{equation}
\Xi = \frac{\eta F_{c}}{cP_0} \ .
\label{eq:Xi1}
\end{equation} 
The phase transition from cold to hot medium occurs with a transcend of
a specific value of the ionization parameter $\Xi$ \citep{BMS83}.
\begin{equation}
\Xi = 0.65 \left(\frac{10^8}{T_e} \right) ^{3/2} \ .
\label{eq:Xi2}
\end{equation}
It scales with the electron temperature and the fraction $0.65$
is the value of the ionization parameter for an inverse Compton temperature
of $10^8 \, \mathrm{K}$.

The set of equations (\ref{Omega}) -- (\ref{eq:Xi2}) allows to
calculate the properties of the corona at each radius, as a function
of global model parameters, i.e. mass of the central black hole, $M$,
the accretion rate, $\dot M$, and the value of viscosity parameter,
$\alpha$, in the corona including the electron temperature and the
optical depth.

The local spectrum is now computed as a disk spectrum
(Eq.~\ref{eq:fnusoft}) Comptonized by the local corona.
Comptonization process was calculated using code based on the
semi-analytical formulae of \citet{CzZ91} for a slab geometry. This
description allows for fast computations on a fine radial grid.

\subsubsection{Global disk/corona spectrum}

The final spectrum is calculated through the integration of the local spectra
over all disk radii. Since the effect of corona is strongly dependent
on the radius - it is weak close to the black hole and reaches maximum
at a distance about a few hundred $\mathrm{R_{Schw}}$, depending on the accretion rate,
this integration is essential. It also means that hard X-ray emission comes 
mostly from the outer parts while weakest Comptonization in the innermost part
creates a soft X-ray tail to the local disk spectrum.  

\subsubsection{Additional Comptonizing cloud}

Our corona model is universal in a sense that it does not contain any
differentiation between radio loud and radio quiet objects while the
composite spectra for the two classes of objects is clearly different.

To model such systematic difference we eventually allow for the
existence of an additional extended medium (see~Fig.~2) with
parameters not determined by the accretion rate.  This medium may be
connected with the observed warm absorber, as a hot medium confining
cooler absorbing clouds \citep{Ogle2000}.  This cloud has a
spherical geometry but we consider also a slab geometry.  Physically
it is characterized by the Thomson opacity, $\tau_{ext}$, and the
electrons temperature, $T_{ext}$. The corona and mainly the disk are
the source of soft photons, which are scaterred by the electrons in
extended medium. In this case they are visible outward as the
additional X-ray radiation.

\begin{figure}[t]
\centerline{
\includegraphics[angle=0.0, height=0.3\textwidth]{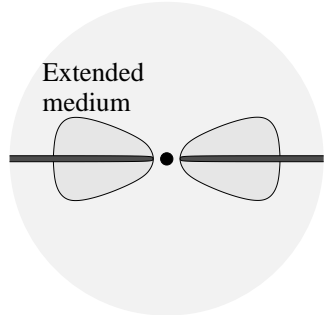}}
\caption{\small A schematic diagram of place the additional Comptonize cloud
denote as the extended medium. This cloud has a spherical geometry. 
It's described by the Thomson scattering $\tau_{Th}$ and the temperature 
of electrons, $T_{em}$, this cloud.
}
\label{mcschem.eps}
\end{figure} 

In this case the Comptonization process is computed once per model so we
use an accurate Monte Carlo method.
We applied the numerical program, which was written by
\citet{PhDGierl} and which is based on procedures by \citet{GorWilcz84}.

Such an accurate method would be too time consuming to apply to the
Comptonization by the corona at each radius and it was not really necessary
since the final integration over the disk surface (an a range of temperatures
and optical depths) smeared any specific Comptonization features even if
present in the spectra (mostly first scattering effect).

\subsubsection{Hard X-ray radiation}

It is generally believed that the hard X-ray radiation in radio-loud objects 
comes from the jet. The extension of the hard X-ray power law into high 
energies is not well constrained in quasars so the nature of this emission 
is not known. The studies of galactic sources in their soft states,
however, suggest that in such a spectral state of an accreting black hole the
hard X-ray emission comes from the non-thermal tail of electron distribution
\citep[e.g.][for \astrobj{Cyg X-1}]{G99}.

Therefore, we do not expect that our thermal corona model will explain the
hard X-ray radiation. We treat this hard X-ray power law emission as an
additional component and we add it to the resulting spectrum adopting the
slope and normalization appropriate for radio loud and radio quiet objects
separately.
  
\section{Results}
\subsection{Without extended medium}
\label{sect:resnocor}

We first discuss the basic version of the model, i.e. disk plus corona,
which is described by the mass of the black hole, $M$, accretion rate, 
$\dot M$, coronal viscosity parameter, $\alpha$, and two parameters 
determining the color temperature distribution in the disk:  $f_{max}$
and $r_{col}$.

The UV part of the composite spectrum is mostly determined by the
first two parameters which govern the total luminosity and the position
of the peak on $\nu F_{\nu}$ diagram. However, the last two parameters are
also important since the multicolor blackbody is too narrow to disk 
represent the \citet{Laor97} spectrum adequately.

\begin{figure}[t]
\centerline{\includegraphics[angle=0, width=0.5\textwidth]{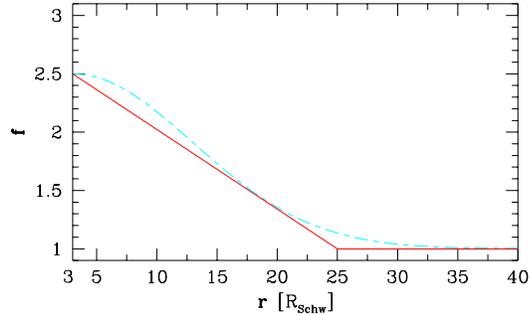}}
\caption{\small The hardening factor $f$ as a function of the
disk radius $r$. 
}
\label{fR.eps}
\end{figure}

However, if we allow for the presence of the saturated Comptonization skin
in the upper layers of the disk we can represent the spectra adequately.

We describe the radial dependence of the color temperature to the 
effective temperature ratio through a linear function of value $f_{max}$
at the marginally stable orbit, dropping to 1 at $r_{col}$ 
(see Fig.~\ref{fR.eps}). 

\begin{figure}[t]
\centerline{\includegraphics[angle=0, width=0.5\textwidth]{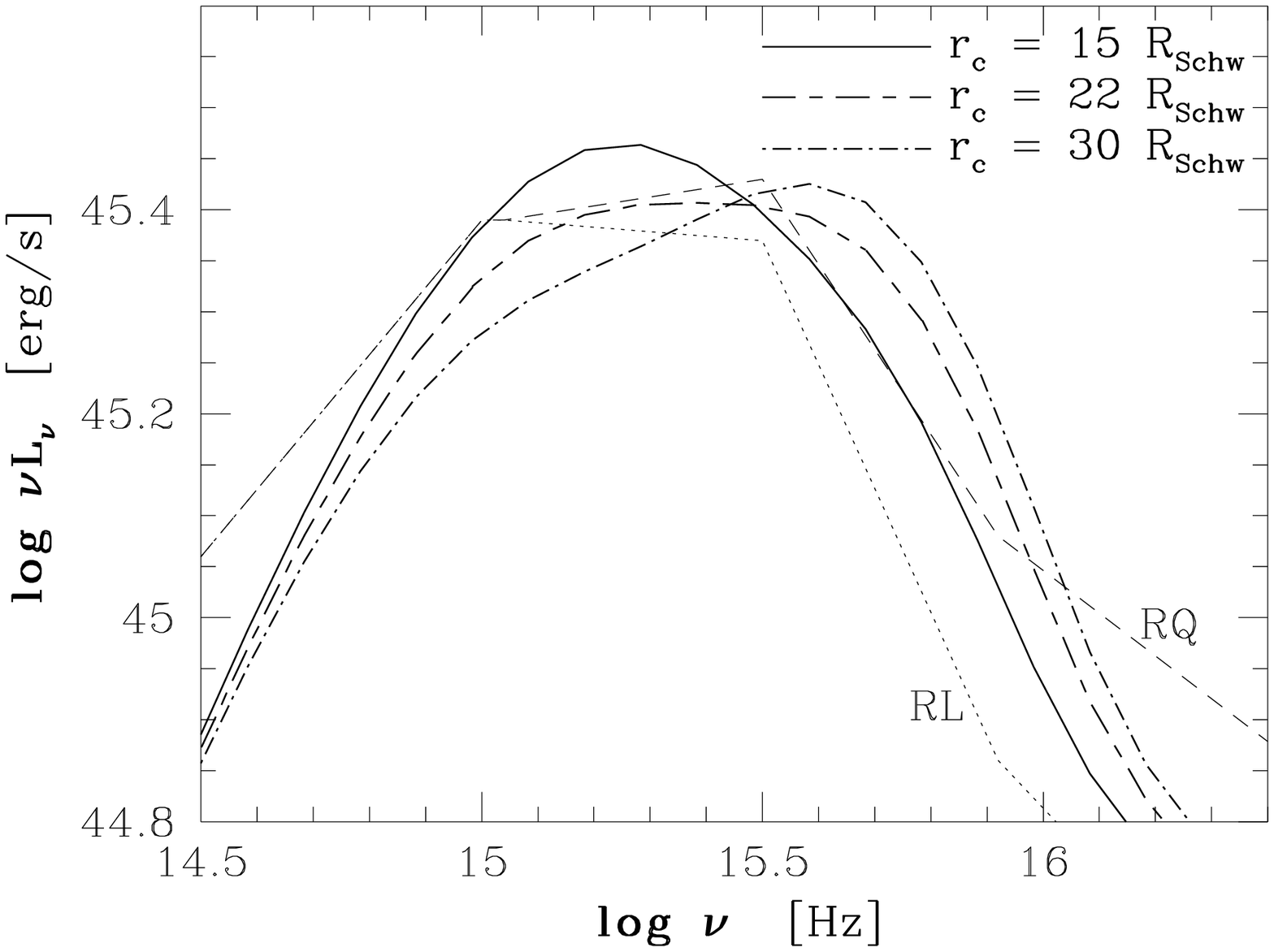}}
\centerline{\includegraphics[angle=0, width=0.5\textwidth]{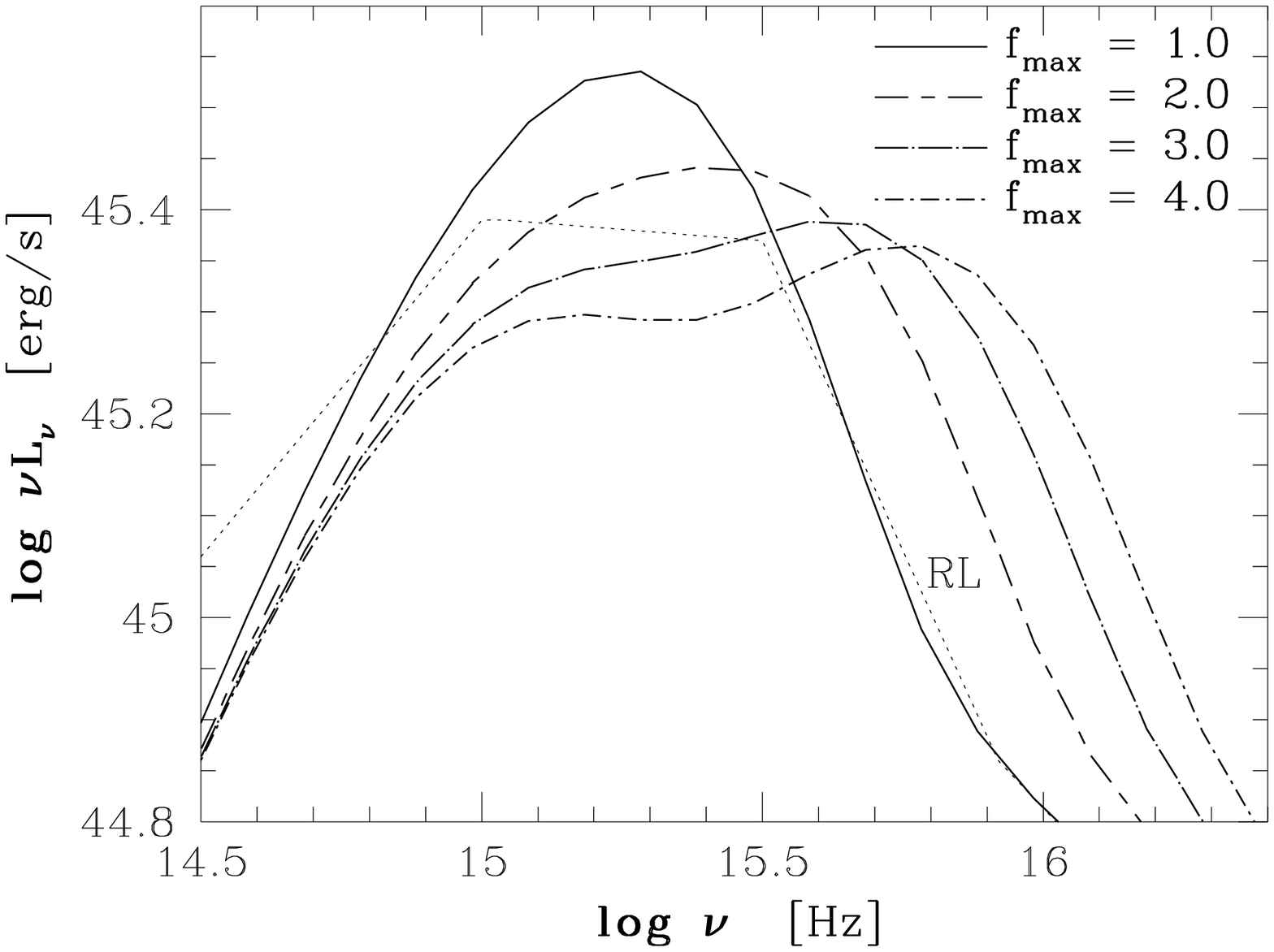}}
\caption{\small The dependence of the spectral shape  
on the adopted value of $r_{col}$ radius (top) for $f_{max}= 2.5$ and
on the value of $f_{max}$ (bottom) for $r_{col} = 25\: \mathrm{R_{Schw}}$. 
Other parameters: mass $M = 1.4 \times 10^9 \: \mathrm{M_{\odot}}$, accretion
rate $\dot M = 1.5 \: \mathrm{M_{\odot}\, yr^{-1}}$, viscosity parameter
$\alpha=0.08$.  To guide the eye, we also show a composite spectrum for
radio loud and radio quiet objects from \citet{Laor97} .  }
\label{varrkf.eps}
\end{figure}

Both parameters, $f_{max}$ and $r_{col}$ broaden the spectral distribution 
in the far UV range but the deviation of the shape is somewhat different in 
those two cases. In Fig.~\ref{varrkf.eps} we show the dependence of the spectrum on
the maximum value of the color to effective temperature ratio, 
$f_{max}$, and on the extension of the disk skin, $r_{col}$ (upper and lower
part, correspondingly). We also experimented with other functional dependences
of the parameter $f$ on the disk radius (see Fig.~\ref{fR.eps}) but when the parameters
were adjusted to model the data possibly well the differences in the final
spectral shapes obtained with different $f(r)$ profiles were negligible.
Therefore, in further discussion we use always the linear formula.

The value of the viscosity parameter $\alpha$ in the corona has no influence
on the optical/UV part of the spectrum but it is important for the relative
efficiency of the X-ray radiation. We choose it to reproduce the observed
level of the soft X-ray emission in the \citet{Laor97} spectrum possibly
well.

\begin{figure}[t]
\centerline{\includegraphics[angle=0, width=0.5\textwidth]{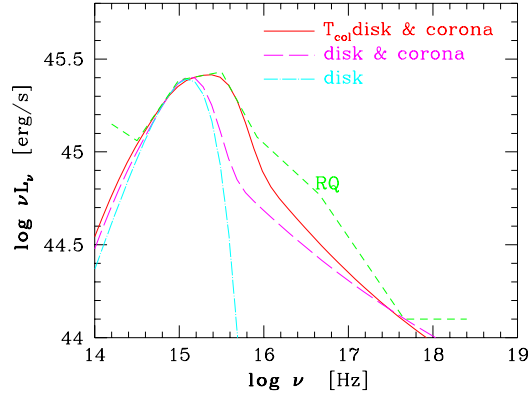}}
\caption{\small The best representation of the composite spectrum of radio 
quiet objects from \citet{Laor97} (short-dashed line) 
with the disk/corona model (continuous line). Model parameters:  
$M = 2.4 \times 10^9 \: \mathrm{M_{\odot}}$,  $\dot M = 1.83 \: \mathrm{M_{\odot}\, yr^{-1}}$, 
$\alpha=0.17$, $f_{max} \simeq 2.3$, $r_{col} = 23\: \mathrm{R_{Schw}}$. For comparison, 
we show the disk blackbody spectrum (dashed-dotted line) and 
disk blackbody+corona spectrum (long dashed line). 
}
\label{rqlts.eps}
\end{figure}

The best representation of the observed composite spectrum for radio quiet
objects is shown in Fig.~\ref{rqlts.eps}. The representation of the data
is satisfactory in the optical/UV band in the X-ray band, where both the
normalization and the slope are well reproduced. The required value of the 
$f_{max}$ parameter implies the color to temperature ratio equal 1.85 at 
$10\; \mathrm{R_{Schw}}$.

However, there is a large
discrepancy between the predicted shape and the observed shape in the
soft X-ray band.

In our model the soft X-ray emission comes from the Comptonization of
the disk emission in the innermost part of the flow. Clearly, the
coronal strength in this region (i.e. the value of the Compton
parameter determined by the temperature and the optical depth) is too
low to model well the observations.  Since our basic model does not
contain additional free parameters apart from the global parameters
listed at the beginning of Sect.~\ref{sect:resnocor} we have no way to
increase this effect artificially. It means that additional
Comptonization is required, possibly by the extended hot medium.

\begin{figure}[t]
\centerline{\includegraphics[angle=0, width=0.5\textwidth]{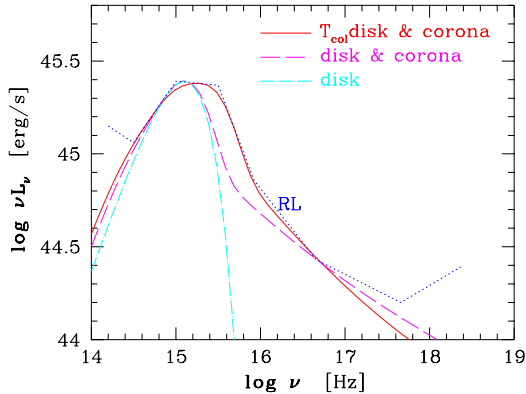}}
\caption{\small The best representation of the composite spectrum of radio 
loud objects from \citet{Laor97} (short-dashed line) 
with the disk/corona model (continuous line). Model parameters:  
$M = 2.9 \times 10^9 \: \mathrm{M_{\odot}}$,  $\dot M = 1.68 \: \mathrm{M_{\odot} \, yr^{-1}}$, 
$\alpha=0.2$, $f_{max} = 2.2$, $r_{col} = 22 \: \mathrm{R_{Schw}}$. 
For comparison, we show the disk blackbody spectrum (dashed-dotted line) 
and disk blackbody+corona spectrum (long dashed line)
}
\label{rllts.eps}
\end{figure}

However, the mean quasar spectrum of radio loud objects in the soft
X-ray band is well reproduced by our basic model (see
Fig.~\ref{rllts.eps}). It means that in this case {\it no extended
medium} is needed to explain the broad band spectrum.

\subsection{Extended medium in radio quiet objects}

The model consisting of the disk and corona emission as described in 
Sect.~\ref{sect:model} did not account for the observed shape of the composite
spectrum of radio quiet quasars 
in the soft X-ray band. Now we consider a case when the disk/corona
system is embedded in an extended hot optically thin cloud.

We fix the parameters of the disk/corona system at the previous values
appropriate for radio quiet and radio loud objects, and we add two new
parameters: the temperature of the extended medium, $T_{ext}$, 
and its optical depth, $\tau_{ext}$.

The soft X-ray emission in the composite spectrum of \citet{Laor97}
is not exactly of a power law shape but shows a discontinuity in the
slope at $log\, \nu \sim 16.7$. Therefore, if we follow this shape
precisely we can determine both parameters of the extended medium
independently. This would not be the case if the emission is a single
power law since in such case only the slope, and subsequently the combination
of the two parameters, is determined. It is difficult to say if this
change of slope is real - it is caused by bridging the unobserved gap
with a power law. However, the values of $T_{ext}$ and $\tau_{ext}$ still
may be quite representative of the required medium.

\begin{figure}[t]
\centerline{\includegraphics[angle=0, width=0.5\textwidth]{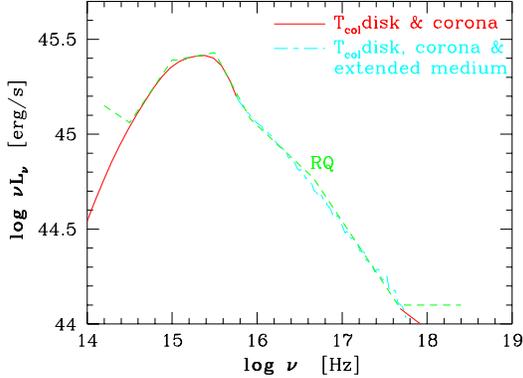}}
\caption{\small The best representation of the composite spectrum of radio 
quiet objects from \citet{Laor97} (short-dashed line) 
with the disk/corona model plus extended medium (continuous line). 
Model parameters:  
$M = 2.4 \times 10^9 \: \mathrm{M_{\odot}}$,  $\dot M = 1.83 \: \mathrm{M_{\odot}\, yr^{-1}}$, 
$\alpha=0.17$, $f_{max} \simeq 2.3$, $r_{col} = 23 \: \mathrm{R_{Schw}}$, 
$T_{ext} = 60-75$ keV, $\tau_{ext} = 0.55-0.5$. 
}
\label{rqcomp.eps}
\end{figure}

The result of this modeling for radio quiet objects is shown in 
Fig.~\ref{rqcomp.eps}. Not surprisingly, this model well represents the
overall spectrum. The hot medium parameters: $T_{ext} = 60-75$ keV and
$\tau_{ext} = 0.55-0.5$, respectively. 

The exact values of the extended medium 
parameters depend on the adopted geometry.
For a comparison, we also considered a flat configuration for the extended
medium. The quality of the data representation is the same.  
The best values of the $T_{ext}$ and $\tau_{ext}$ are equal $55$ keV 
and $0.45$.

\section{Discussion}

We considered a physically justified model of the disk/corona. The emission
from such a system calculated in the simplest direct way did not account 
for the extended shape of the spectrum in far UV and for the shape of the 
soft X-ray spectrum in
radio quiet and radio loud objects in the composite spectra of \citet{Laor97}.

Two modifications to the simplest picture were necessary: (1) the presence of 
the hotter disk skin where saturated Comptonization leads to a noticeable
difference between the color temperature and the effective temperature
at $r < 30\; \mathrm{R_{Schw}}$. (2) the presence of the hot extended medium surrounding
the disk and the corona in the case of radio-quiet objects. 

\subsection{The reliability of the composite spectra}
\label{sect:other}

Our results depend sensitively on the spectral shape of the composite
spectra adopted as the representation of a typical spectrum of radio loud and 
radio quiet quasar, respectively.

The reliability of the composite spectrum of \citet{Laor97} 
may be easily questioned since various types of objects were used to
construct two major parts: high redshift quasars ($\mathtt{z} \sim 1$) were used
by Zheng et al. (1997) to obtain the far UV part and low redshift objects
were used to find the soft X-ray part. This was unavoidable since this was 
the only way to close the observational gap between the two bands. Also the
number of objects used to prepare the spectrum ($\sim 100$) was not very large.

The confidence in the composite may be further decreased by the presence of
systematic differences between this composite and other composites in the
optical/UV band.

\begin{figure}[]
\centerline{\includegraphics[angle=0, width=0.5\textwidth]{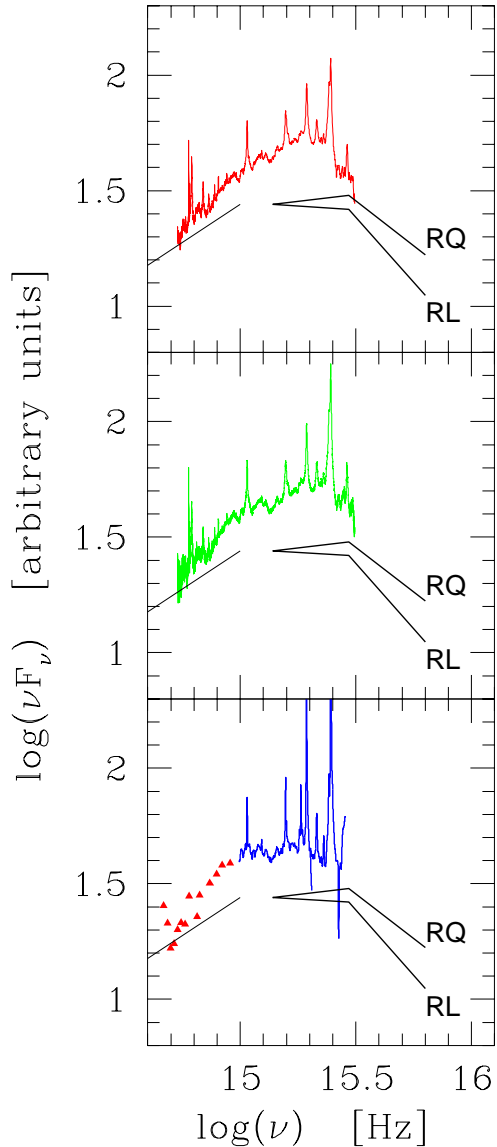}}
\caption{\small The comparison of the Zheng et al. (1997) composite in the 
UV/far UV band (thick solid line) and Laor et al. (1997) in the
optical band (thin solid line) with Francis et al. (1991) composite
(upper panel), Morris (private communication) composite (middle panel)
and the spectrum of a quasar/NLS1 PG1211 from Elvis et al. (1994,
optical data) and Bechtold et al. (2001, HST data).  
}
\label{fig:kompozyty}
\end{figure}

In Fig.~\ref{fig:kompozyty} we show the comparison of schematically
drawn Zheng et al. (1997) composite and the optical part of the Laor
et al. (1997) against the Francis et al. (1991; upper panel) and
Morris (private communication; middle panel). We see strong
flattening in Zheng et al. which lead to the conclusion about high
color to effective temperature ratio in quasar accretion disks. Almost
no such flattening is present in two other composites: standard black
body accretion disk is almost a perfect representation of the data for
Francis et al. \citep[see Fig.~17 in the review of][]{KB99}.

The difference, however, is most probably due to the difference in the
studied population of quasars. Francis et al. (1991) sample consists
of quasars at $\mathtt{z} \sim 3$ and of the bolometric luminosity of
order of $10^{47} - 10^{48}\; \mathrm{erg\: s^{-1}}$ and Morris
composite is based on the same sample while \citet{Zheng97} part of
the Laor composite was based on $\mathtt{z} \sim 1$ quasars of
bolometric luminosity of order of $10^{46}\; \mathrm{erg\: s^{-1}}$.

As an argument in favor of this interpretation we plot in
Fig.~\ref{fig:kompozyty} the optical/UV spectrum of PG1211+143 - a radio
quiet quasar classified also as NLS1 galaxy. The bolometric luminosity
of this object is $2 \times 10^{45}\; \mathrm{erg\: s^{-1}}$ 
\citep{JCzM01}. The shape of this spectrum is noticeably similar to
the composite and cannot be modeled with a black body accretion disk.
It is definitely flatter in UV than \citet{Fra91} composite.
On the other hand, individual spectra of high redshift, bright quasars
show on average much harder spectra, with mean value of the slope of
0.36 (i.e. $F_{\nu} \propto \nu^{-0.36}$) for a total of 24 radio quiet 
quasars, consistent with
\citet{Fra91}, in the observed 1000 - 2000 \AA ~ band in quasar
rest frame \citep{Scott2000}. 

The dependence of the spectral shape on the luminosity state is well
known for galactic sources in their high/soft states dominated by the
disk emission.  Sometimes the soft component is well described as disk
black body emission but frequently additional Comptonization of this
component is needed, independent from the hard X-ray tail \citep[for a
review see e.g.][]{Zycki01}.  We therefore conclude that our results
apply to faint or moderately bright quasars, and do not apply to
bright quasars.

Unfortunately, we cannot repeat the whole analysis for bright objects.
The composite of \citet{Fra91} appropriate for bright quasars is not
supplemented by the X-ray part. The data on single bright objects
suggest that the hard X-ray slope \citep{ReevTur2000} is similar to that
given by Laor et al. (photon index $\Gamma = 1.89 \pm 0.05$ for radio
quiet quasars and $1.66 \pm 0.04$ for radio loud quasars). The
relative level of hard X-ray emission in comparison to UV in very
bright objects may also be similar: for example, very bright quasar PG
1247+268 has $log\, \nu L_{\nu}$ equal to 47.09 at 3000 \AA~ 
\citep{Neugeb87} and 45.72 at 2 keV \citep{ReevTur2000}, 
meaning the same $\alpha_{ox}$ index as determined by \citet{Laor97}.
However, the soft X-ray part becomes unconstrained since
the soft X-ray excess moved into unobserved range. Therefore, it
cannot be used effectively to test the corona model.

\subsection{Mean quasar properties}
 
Our disk/corona model implies that the studied quasar sample does not
accrete at a rate close to the Eddington limit. The value of the mass of the 
representative black hole is about $3 \times 10^9\: \mathrm{M_{\odot}}$ for radio
loud objects and somewhat lower for radio quiet objects ($\sim 2.5 \times 
10^9\: \mathrm{M_{\odot}}$), and the luminosity to the Eddington luminosity ratio
is equal 0.016 and 0.022, correspondingly, for a non-rotating black hole
and pure hydrogen plasma. This ratio is rather low, typical for Seyfert 1
galaxies, only the mass of the black hole and consequently, the luminosity is
proportionally higher. This result is a direct consequence of the high
color to effective temperature ratio since roughly
\begin{equation}
L/L_{Edd} \propto f^{-2}\ .
\end{equation}
Disk blackbody fits would therefore indicate the ratios 3-4 times higher but
we did not find them satisfactory in case of \citet{Laor97} composite.
Our value of the black hole mass agrees with the value obtained by Kawaguchi 
et al. (2000) since the model contained similar color to effective 
temperature ratio. Their accretion rate for radio quiet quasars is slightly 
higher ($L/L_{Edd} = 0.031$) because their model effectively accounted also
for the soft X-ray excess, although at the expense of arbitrary assumptions
about the corona properties. 

Fits to Francis et al. (1991) composite would give similar value of the
black hole mass \citep[see e.g.][]{KB99} but much higher accretion 
rate, and the $L/L_{Edd}$ ratio closer to 1.

The value of the viscosity parameter determined from the model ($\alpha \sim 
0.2$) is reasonable. Similar values are derived for accretion disks in 
cataclysmic variables (e.g. Smak, 1999) and in galactic X-ray novae during
outburst (e.g. Lasota, 2001). Models of fast variability of the microquasar
GRS 1915+105 under the influence of radiation pressure instability suggest
somewhat smaller value, $\alpha \sim 0.01$ (Janiuk, Czerny \& Siemiginowska, 
2000). Determinations available for AGN based on the observed optical 
variability supposed to correspond to thermal timescale give rather weak
constraints ($\alpha > 0.01$, Siemiginowska \& Czerny, 1989; $\alpha \sim 0.03$,
Webb \& Malkan, 2000). 

The value of the viscosity parameter is essential from the point of
view of the inner boundary condition of an accretion disk. Low values,
below 0.1, are consistent with the standard description of the
accretion in the vicinity of the marginally stable orbit while $\alpha
>0.1$ leads to a considerable loss of angular momentum below the
marginally stable orbit due to the type of the sonic point
\citep{M-Cz86,KFM98} or strong magnetic field (Krolik, 1999; Reynolds,
Armitage \& Chiang, 2001). Determined values are just at the border.

\subsection{Disk skin}

Large color to temperature ratio required by the data is an interesting
issue since it requires low level of absorption and strong dominance
of the disk atmosphere by electron scattering.
 
Significant ionization of the disk surface at small radii is not
unexpected.  The observation of the X-ray reflected component is a
complementary way to estimate the disk opacity close to the surface
observationally (for a review, see Mushotzky, Done \& Pounds,
1993). Although the study of quasars are much more difficult than the
study of Seyfert galaxies Nandra et al. (1997) and recently Reeves \&
Turner (2000) were able to determine the iron $K_{\alpha}$ line
properties on the basis of ASCA data. The line energy is at $\sim
6.57$ keV for objects of the brightness comparable to those in the
composite spectrum, which corresponds to the value of the ionization
parameter $\xi$ about a few hundred (e.g. \. Zycki \& Czerny,
1994). The medium at the temperature $\sim 10^6$ K and optical depth
$\tau_{es} \sim 1$ is expected to develop under moderate external
irradiation by the (non-thermal?) hard X-ray component since the
temperature corresponds to an Inverse Compton temperature for
effectively soft total radiation field \citep[e.g.][]{CzD98,Nayakshin2000}.

However, such a skin will result in very weak Comptonization effect, and in 
particular, it will not change the color temperature of the Big Blue Bump peak.
To explain the color to effective temperature ratio of about 2 we need a thick
layer ($\tau_{es} \sim 100$) remaining at the color temperature due to some
heating and due to cooling by saturate Comptonization. The absorption in this
zone should be negligible - otherwise the temperature would drop down to the
effective temperature.

The results of \citet{ShiTa95} as well as \citet{KawaShiMi2000}
predict such a strong Comptonization because the computations
do not contain bound-bound and bound-free transitions which are efficient
coolants at the discussed temperatures, particularly oxygen, carbon and 
iron lines are important. Computations of irradiated media including those
transitions do not show the color to effective temperature ratio in the 
thermalised emission significantly different from 1 
(e.g. Madej \& R\' o\. za\' nska, 2000; Nayakshin et al., 2000; Dumont et al., 
2001). The required level of ionization for efficient Comptonization seems to
be higher than naturally predicted in an irradiated medium. Similar problem
was found by \citet{CzeZ94} when they tried to explain the
universal position of the soft X-ray excess as the reflection feature from the
weakly ionized medium. The presence of the heavy corona further suppresses
the ionization level at the disk surface if the corona is in the hydrostatic
equilibrium with the disk \citep{Roza99}.

It is possible that the effect is caused by significant generation of the
energy close to the disk surface. Such an enhanced generation was not
included in the performed radiative transfer computations. It would mean
that the vertical distribution of heat generation is not well parameterized
by the $\alpha P$ prescription and the magnetic field may be responsible for 
such a global redistribution.

On the other hand, the conclusion on the large color to effective temperature 
ratio relies strongly on the shape of the composite spectrum in UV -- far UV
band which still has to be tested (see Sect.~\ref{sect:other}).

\subsection{Soft X-ray discrepancy for radio-quiet objects}

The adopted corona model adequately described the soft X-ray part of the 
composite spectrum for radio loud objects. Low optical depth of the corona
and a broad electron temperature range predicted by the model at different 
radii was adequate to give the observed effect of moderate Comptonization.

However, in this model the Comptonization is definitely too weak to
account for the soft X-ray spectrum of radio quiet objects.  Similar
modeling of the composite spectrum of radio quiet objects was
performed by \citet{KawaShiMi2000} and in their case
the modeling was successful. This was caused by the fact that in their
approach the fraction of the energy dissipated in the corona and the
optical depth of the corona at a given radius were free parameters of
their model. In our corona model the radial distribution of those
quantities were calculated from the global parameters (black hole
mass, accretion rate and viscosity parameter) since the spontaneous
division of the flow into the disk and corona part results from the
cooling properties of the material in the optically thin and optically
thick part. As a result, our corona was never as thick as assumed by
Kawaguchi et al., independently from the assumed viscosity parameter.

Our corona model, adopted after Janiuk \& Czerny (2000), contained
several simplifications. However, recent progress in corona modeling
does not seem to improve the situation. If the corona is assumed to be
accreting, and the advection is included, the corona above a disk is
even weaker than in our description \citep{JZCz2000}.  Better
description of the disk/corona transition and replacement of
Eqs.~\ref{eq:Xi1}--\ref{eq:Xi2} with the disk evaporation condition
due to the electron conduction (R\' o\. za\' nska \& Czerny, 2000a)
leads again to the prediction of still weaker corona for high
accretion rates (R\' o\. za\' nska \& Czerny, 2000b). On the other
hand, the presence of the magnetic field may weaken the efficiency of
the conduction and reduce the condensation effect found in the
innermost part of the disk accreting at a high rate.

Although a theory did not say its last word on this subject as the
formation of the corona is a badly understood process it may suggest
that actually the disk corona is too weak to produce the soft X-ray
spectrum of radio quiet quasars and the existence of the additional
Comptonizing medium may be necessary.

\begin{figure}[t]
\centerline{\includegraphics[angle=0, width=0.5\textwidth]{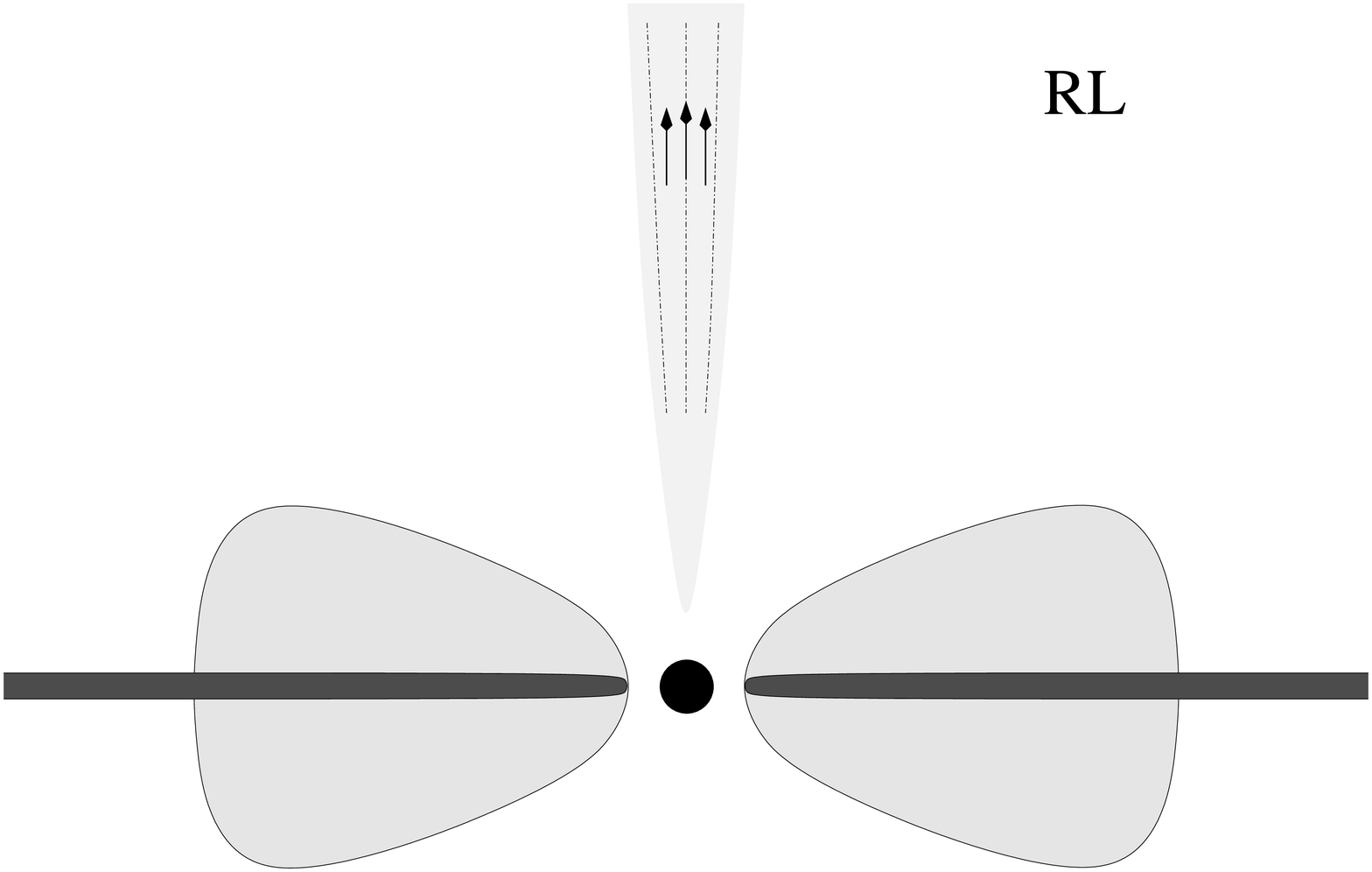}}
\centerline{\includegraphics[angle=0, width=0.5\textwidth]{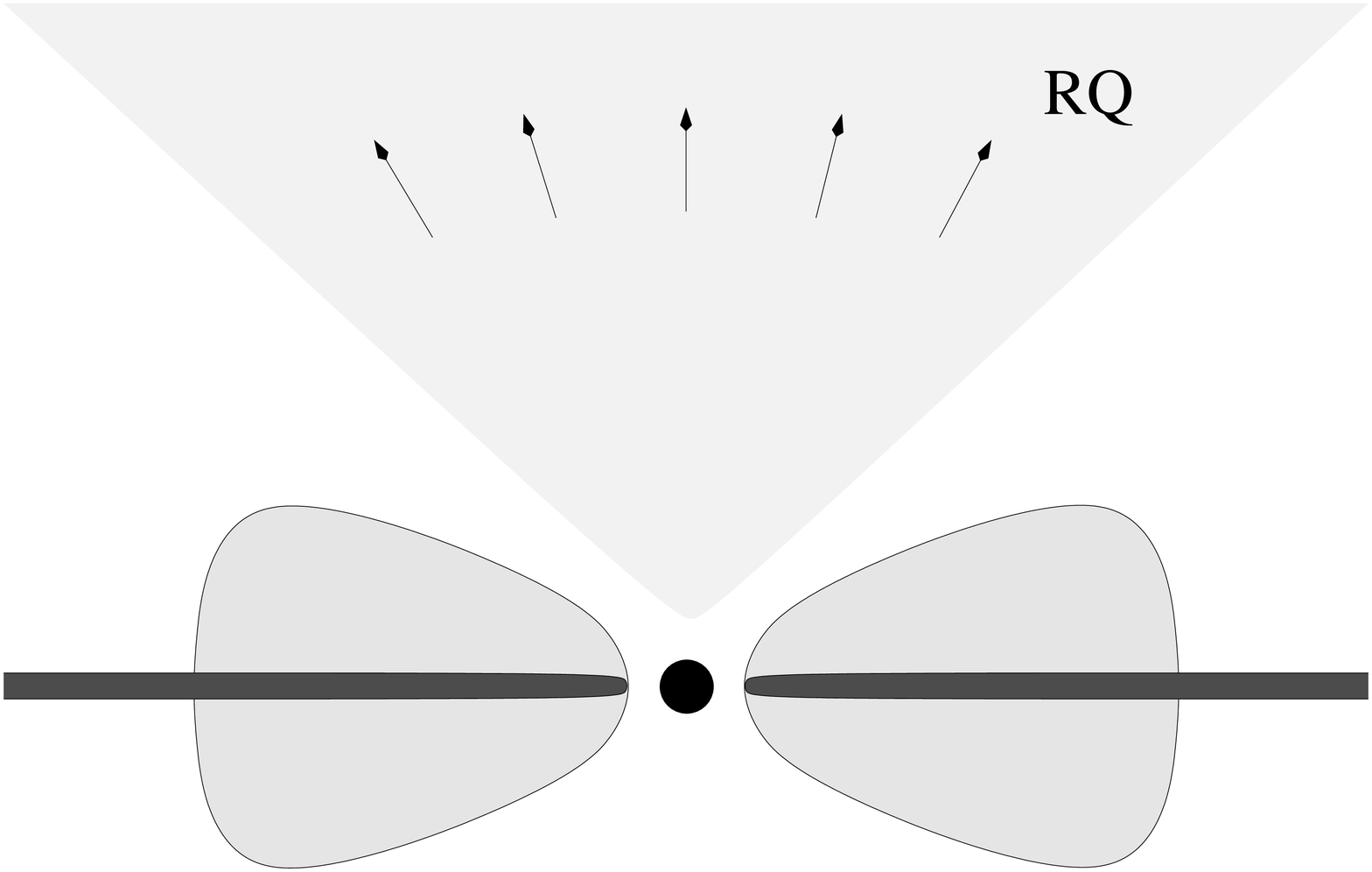}}
\caption{\small Possible character of an outflow in radio loud (RL)
and radio quiet (RQ) quasars. 
}
\label{outflow}
\end{figure}

The comparison of the radio loud objects and radio quiet objects may
suggest that the outflow takes place in all types of AGN
\citep[for a review, see e.g.][]{Veilleux2000}. In radio loud objects the
outflow is collimated in a form of jet, so the jet contributes to the
hard X-ray emission but generally it does not cover our line of sight
to the most part of the accretion flow.  The mechanism of this
collimation and its distance from the black hole is still under debate
(e.g. Sikora, 2001).  In radio quiet objects the outflow perhaps is not
collimated so the outflowing plasma is in our line of sight and
Comptonizes the disk/corona emission.  We show such a putative
geometry in Fig.~\ref{outflow}.

Some kind of outflow is frequently predicted by the corona models or
two-dimensional computations of the accretion flow 
\citep[e.g.][]{BMS83, MM-H94, Igumensh2000, TurDull2000}.
However, the estimate of the
amount of outflowing mass and its physical conditions is difficult and
no predictions differentiating between the radio loud and radio quiet
objects are available within the frame of this approach. The
properties of the outflowing material may significantly depend on the
considered distance from the black hole as well as the inclination
angle which further complicates the issue (e.g. Elvis, 2000).

\subsection{Observational tests - soft X-ray emission lines}

The dominant role of the Comptonization in formation of the soft X-ray
spectra implies that no strong emission lines are expected in this
band. On the other hand, models of coronal flares predict that a
significant fraction of the disk is not covered by the corona but it
is strongly irradiated. Such irradiated internally not heated layers
are the source of rich emission line spectrum in soft X-ray band
\citep[e.g.][]{Nayakshin2000} since the hot layer is
optically thin due to thermal instability in the heated/cooled plasma
(Krolik et al., 1981; R\' o\. za\' nska \& Czerny, 1996).

High resolution observations were performed so far in the case of a
few Seyfert 1 galaxies, and the results were not definite:
Branduardi-Raymond et al., (2000) claimed that broad oxygen, nitrogen
and carbon emission lines, of similar shape to $K_{\alpha}$ iron line,
are present in the XMM-Newton spectra of MCG-6-30-15 and Mrk 766 while
Lee et al. (2001) argue that they see only the warm absorber effect on
the continuum in the Chandra spectrum of MCG-6-30-15. More detailed
analysis of new XMM and Chandra data should resolve this issue.

\section{Conclusions}

\begin{itemize}

\item disk/corona model of Janiuk \& Czerny (2000) based on thermal 
instability well describes the composite spectrum of radio loud faint
quasars in the optical-soft X-ray band if the color to effective
temperature ratio is close to 2 in the innermost part of the disk

\item adopted disk/corona model does not account for the soft X-ray spectrum
in radio quiet faint quasars and additional Comptonizing medium
is required, possibly due to the uncollimated outflowing hot plasma

\end{itemize}

\section*{Acknowledgements}

We thank Paul Francis and Simon Morris for providing us with their
composite spectra and Jill Bechtold and Adam Dobrzycki for the HST
data of PG1211 used in Fig.~\ref{fig:kompozyty}. We thank also Marek
Gierli\'nski for his computer program and helpful discussion.

This work was supported in part by grant 2P03D01816
of the Polish State Committee for Scientific Research.

M. Nikolajuk would like to thank Prof. Dr T. Trojanowski and Dr E. Maziarz from the Neuro surgical 
Wards of the Clinical Hospital in Lublin and dedicate this paper to P. Sarzala (coma) and to memory of M. Kalicki (cancer).

\end{document}